\documentclass[aps,prb,twocolumn,groupedaddress,citeautoscript,longbibliography]{revtex4-1}
\usepackage[dvips]{graphicx}
\usepackage{amsmath}
\usepackage{amssymb}
\usepackage{color,colortbl}
\usepackage{bm}

\newcommand{\tm}[1]{\textrm{#1}}

\newcommand{\avg}[1]{\langle#1\rangle}
\newcommand{\ks}{\bm{k}s}

\newcommand{\kk}{\bm{k}}
\newcommand{\q}{\bm{q}}
\newcommand{\qs}{\bm{q}\sigma}
\newcommand{\qa}{\bm{q}\alpha}
\newcommand{\qap}{\bm{q}'\alpha'}
\definecolor{snow}{rgb}{0.92,0.91,0.91}
\definecolor{lgray}{rgb}{0.83,0.83,0.83}
\definecolor{lyell}{rgb}{1,1,0.88}
\definecolor{gwhite}{rgb}{0.97,0.97,1}

\begin{document}


\title{Chiral Phonons and Electrical Resistivity of Ferromagnetic Metals at Low Temperatures}


\author{E. Solano-Carrillo}
\affiliation{Department of Physics, Columbia University, New York, NY 10027, USA}


\begin{abstract}
Ferromagnetism is an exciting phase of matter exhibiting strongly correlated electron behavior and a standard example of spontaneously broken rotational symmetry: below the Curie temperature, atomic magnets in an isotropic single-domain ferromagnetic metal align along a spontaneously chosen direction. The scattering of conduction electrons from thermal perturbations to this spin order, together with electron-electron collisions, mark the material electrical behavior at low temperatures, where the resistivity varies mostly quadratically with the temperature. Around liquid-helium temperatures however, an interesting phenomenon occurs, giving rise to an extra \emph{linear} contribution to the variation of the electrical resistivity with temperature, whose theoretical explanation has encountered problems for a long time. Here I introduce a spin-flip scattering mechanism of conduction electrons in ferromagnetic metals arising from their interaction with the internal magnetic induction and mediated by chiral modes of the crystal lattice vibrations carrying spin 1. This mechanism is able to explain the above anomaly and give a good account of the spin-lattice relaxation times of iron, cobalt and nickel at room temperatures.
\end{abstract}


\maketitle

\section{Introduction}
For decades, the microscopic process which causes a linear-in-temperature term in the electrical resistivity of pure ferromagnetic metals (Fe, Co and Ni) at low temperatures---which is clearly observed around liquid-helium temperatures \cite{Campbell,Volkenshtein}---has remained unclear. In this temperature region, the $T^2$ dependence of the electrical resistivity characteristic of the transition metals at low temperatures, due to the $s$-$d$ exchange interaction \cite{Kasuya2,Goodings,Mannari} and inter-electronic collisions, \cite{Baber} ceases to be the only dominant contribution.  The most known intrinsic mechanism giving a linear term in the resistivity is the spin-orbit interaction between the orbits of the $4s$ conduction electrons and the spins of the nearly localized $3d$ ferromagnetic electrons. \cite{Turov,Turov2,Turov3} However, this predicts a linear coefficient which is about a thousand of times smaller than observed. \cite{Turov2,Goodings,Taylor}

Despite other mechanisms have been proposed \cite{Volkenshtein} to explain this anomalous behavior, including e.g. electron-magnon scattering taking into account the electronic spin polarization, and scattering of the conduction electrons by 2D spin-wave excitations on the magnetic domain walls; it is believed, \cite{Campbell,Volkenshtein} based on a series of experiments, that the anomaly is caused by the scattering of conduction electrons by the internal magnetic induction present in the ferromagnetic metals, observed as an internal magnetoresistance effect. However, no explanation of this fact has been given so far using quantum mechanics.

In this article, I propose a simple picture of an internal magnetoresistance effect in the ferromagnetic metals which predicts the correct magnitude of the linear coefficient. This is realized as the contribution to the electrical resistivity coming from electronic spin-flip transitions in the conduction band---which is Zeeman-split by the internal magnetic induction---and mediated by the isotropic spin-phonon interaction of the conduction-electron spins with the orbital \emph{contact} (hyperfine) field these electrons produce at the ionic positions. This mechanism, which accounts for the observed spin-lattice relaxation times of pure ferromagnetic metals at room temperatures, complements the existing theories of spin relaxation of conduction electrons in metals, \cite{Overhauser,Fabian,Boross,Mokrousov} which do not deal with the ferromagnetic case.

The electronic spin-flip transitions introduced here portrait phonons as carriers of angular momentum. The macroscopic consequences of this were first discussed by Zhang and Niu\cite{Zhang} in their consideration of the Einstein-de Hass effect in a magnetic crystal, leading to the envisioning of \emph{chiral} phonons\cite{ZhangNiu2} as lattice modes supporting left-handed and right-handed excitations and spin\cite{Garanin,Holanda}. The direct observation of chiral phonons has been done very recently\cite{SKim,HZhu}; however, as far as I know, the role played by them in the electrical resistivity of a metal has not been considered before.    

\section{Description of the model}

Consider a system of itinerant and interacting electrons \emph{magnetically} coupled to the localized ions of the material. The Hamiltonian of this system is
\begin{equation}\label{Hked}
 H_e = \sum_{\kk s}E_k n_{\kk s}+\dfrac{1}{2}\sum_{\kk \neq 0}J(\kk)\rho_{\kk}\rho_{-\kk}+H_{\tm{dd}}.
\end{equation}
Here, the first term represents the kinetic energy of these electrons, which have wave number $\kk$ and spin index $s$, with $E_k=\hbar^2k^2/2m$ and $n_{\kk s}=\hat{c}_{\kk s}^{\dagger}\hat{c}_{\kk s}$ being the electron number operator. The second term represents the electron-electron Coulomb interactions, with $\rho_{\kk}=\sum_{\bm{l}s}\hat{c}_{\bm{l}+\kk,s}^{\dagger}\hat{c}_{\bm{l} s}$ being a Fourier component of the electronic density, and $J(\kk)$ being the Fourier transform of the Coulomb electric potential. 

The third term in \eqref{Hked} represents the magnetic dipole-dipole interactions between electron pairs and between electron-ion pairs. This is given by
\begin{equation}\label{dd}
 H_{\tm{dd}}=-\sum_{r}\bm{\mu}_r\cdot\bm{B}(\bm{x}_r)=-\sum_{rq}\bm{\mu}_r\cdot \bm{D}(\bm{x}_r-\bm{x}_q)\cdot\bm{\mu}_q, 
\end{equation}
where $\bm{\mu}_r$ is the magnetic moment of the $r^{\tm{th}}$ dipole at position $\bm{x}_r$, which interacts with the magnetic dipole field $\bm{B}(\bm{x}_r)=\sum_q  \bm{D}(\bm{x}_r-\bm{x}_q)\cdot\bm{\mu}_q$ generated by the other dipoles. Here  $\bm{D}(\bm{x}_r-\bm{x}_q)$ is a dyad representing the dipole kernel 
\begin{equation}\label{dk}
\bm{D}(\bm{x}_r-\bm{x}_q) = \dfrac{3\,\hat{\bm{x}}_{rq}\hat{\bm{x}}_{rq}-\bm{1}}{|\bm{x}_r-\bm{x}_q|^3}+\dfrac{8\pi}{3}\delta(\bm{x}_r-\bm{x}_q)\bm{1},
\end{equation}
with $\hat{\bm{x}}_{rq}$ a unit vector from $\bm{x}_r$ to $\bm{x}_q$---note that the second term in \eqref{dk} is necessary to account for the volume integral of the magnetic dipole field $\bm{B}(\bm{x})$ over a region containing all the dipoles.\cite{Jackson} 

Let me divide now the magnetic dipoles in two classes: those belonging to the ions, in which case the label is changed to $r_i$, and those belonging to itinerant electrons, in which case the label is changed to $r_e$. Furthermore, by separating the orbital and spin contributions to the magnetic moments as
\begin{equation}\label{mur}
 \bm{\mu}_r = \bm{\mu}_r^{\tm{spin}}+\bm{\mu}_r^{\tm{orb}}=-\mu_B(2\bm{S}_r+\bm{L}_r),
\end{equation}
where $\bm{S}_r$ and $\bm{L}_r$ are, respectively, the spin and orbital angular momentum operators (in units of $\hbar$) corresponding to the $r^{\tm{th}}$ dipole---for simplicity assume an electronic $g$-factor of 2---and substituting \eqref{mur} into \eqref{dd}, the following contributions to the dipole-dipole Hamiltonian turn out to be sufficient for the discussion
\begin{equation}\label{Hbi}
\begin{split}
 H_{\tm{\,spin-i}}^{\tm{\,spin-e}}&=-\sum_{r_e}\bm{\mu}_{r_e}^{\tm{spin}}\cdot\bm{B}_i^{\tm{spin}}(\bm{x}_{r_e})\\
 &=-\sum_{r_e,q_i}\bm{\mu}_{r_e}^{\tm{spin}}\cdot \bm{D}(\bm{x}_{r_e}-\bm{x}_{q_i})\cdot\bm{\mu}_{q_i}^{\tm{spin}},
 \end{split}
\end{equation}
which represents the interaction of the spins of the itinerant electrons with the magnetic field created by the spins of the ions;
\begin{equation}\label{Hci}
\begin{split}
 H_{\tm{\,orb-e}}^{\tm{\,spin-e}}&=-\sum_{r_e}\bm{\mu}_{r_e}^{\tm{spin}}\cdot\bm{B}_e^{\tm{orb}}(\bm{x}_{r_e})\\
 &=-\sum_{r_e,q_e}\bm{\mu}_{r_e}^{\tm{spin}}\cdot \bm{D}(\bm{x}_{r_e}-\bm{x}_{q_e})\cdot\bm{\mu}_{q_e}^{\tm{orb}},
 \end{split}
\end{equation}
which represents the interaction of the spins of the itinerant electrons with the magnetic field created by the orbital motion of these electrons; and
\begin{equation}\label{Hoo}
\begin{split}
 H_{\tm{\,orb-i}}^{\tm{\,orb-e}}&=-\sum_{r_e}\bm{\mu}_{r_e}^{\tm{orb}}\cdot\bm{B}_i^{\tm{orb}}(\bm{x}_{r_e})\\
 &=-\sum_{r_e,q_i}\bm{\mu}_{r_e}^{\tm{orb}}\cdot \bm{D}(\bm{x}_{r_e}-\bm{x}_{q_i})\cdot\bm{\mu}_{q_i}^{\tm{orb}},
 \end{split}
\end{equation}
which represents the interaction of the orbital moments of the itinerant electrons with the magnetic field created from these same orbits.

The terms that I have neglected from $H_{\tm{dd}}$ are the spin-orbit interactions $H_{\tm{\,orb-i}}^{\tm{\,spin-e}}$ and $H_{\tm{\,spin-i}}^{\tm{\,orb-e}}$ as well as $H_{\tm{\,orb-e}}^{\tm{\,orb-e}}$ and $H_{\tm{\,spin-e}}^{\tm{\,spin-e}}$, whose contributions to the electrical resistivity of metals are well known \cite{Turov2,Taylor,Overhauser} and therefore do not play an important role in the appearance of the effect here described.

To proceed further, I make the standard assumption that the magnetization of the ferromagnetic body is entirely due to the unbalanced spins of the $3d$ electrons in the ions---the $4s$ electrons being the itinerants---and replace $\bm{B}_i^{\tm{spin}}(\bm{x})$ in \eqref{Hbi} with the average internal magnetic induction $\bm{B}_i(\bm{x})=4\pi \bm{M}(\bm{x})$ in the absence of an externally applied magnetic field, where $\bm{M}(\bm{x})$ is the magnetization field---this is just the magnetic constitutive relation involved in the macroscopic Maxwell's equations. 

At the low temperatures of interest and zero applied fields, the spatial dependence of the internal magnetic induction may be suppresed---corresponding to the neglection of the magnetic domain structure, whose effect was mentioned in the introduction to be irrelevant. As a result
\begin{equation}\label{Bi}
 \bm{B}_{i}=4\pi \bm{M}_s,
\end{equation}
with $\bm{M}_s$ being the magnetization within an arbitrary magnetic domain---typically measured to be the saturation magnetization \cite{Gurevich}---here taken to point along an arbitrary direction consistent with the neglection of magnetic anisotropies. According to de Haas-van Alphen oscillation experiments, \cite{Berger,Anderson,Joseph} Eq. \eqref{Bi} is practically \emph{the} internal magnetic induction seen by conduction electrons in the ferromagnetic metals within each magnetic domain. 

Having approximated the effect of \eqref{Hbi} by introducing the average field \eqref{Bi} acting on the itinerant-electron spins, I turn now to \eqref{Hci}. In the magnetic field $\bm{B}_e^{\tm{orb}}(\bm{x})$ due to the electronic orbital motion, the underlying isotropy of the present model calls for the neglection of magnetic anisotropy effects related to the first term in \eqref{dk}. Consequently, taking the orbital part of \eqref{mur}, I consider only the contact (hyperfine) field due to the itinerant-electron orbital motion
\begin{equation}\label{Borb}
 \bm{B}_e^{\tm{orb}}(\bm{x})=-\dfrac{8\pi}{3}\mu_B\sum_{q_e}\delta(\bm{x}-\bm{x}_{q_e})\,\bm{L}_{q_e}.
\end{equation}
It is in the treatment of this field that the main results of this paper rest upon. This is done next. 
\subsection{Main assumption regarding the electronic orbital motion}
My claim is that the magnetic field contribution \eqref{Borb} from the orbital motion of the itinerant electrons gives a noticeable effect provided these electrons are in a state where they \emph{rigidly} move with the ions---while still being able to drift in the transport of electricity. This is expressed mathematically by writing in \eqref{Borb}
\begin{equation}\label{xL}
 \bm{x}_{q_e}=\bm{R}_n+\bm{r}_n,\hspace{0.5cm} \bm{L}_{q_e}=\bm{L}_{n},
\end{equation}
where $\bm{L}_n$ is the orbital angular momentum (in units of $\hbar$) of the $n^{\tm{th}}$ \emph{ion}, which at a given time is displaced $\bm{r}_n$ from its equilibrium position $\bm{R}_n$. 

For the low temperatures of interest, the displacement $\bm{r}_n$ is very small compared to interatomic distances and then \eqref{Borb} is, to leading order,
\begin{equation}\label{Bc}
\bm{B}_c(\bm{x})=-\dfrac{8\pi}{3}\mu_B\sum_n\delta(\bm{x}-\bm{R}_n)\,\bm{L}_n,
\end{equation}
where the subscript $c$ stands for ``contact'' field which, although it is entirely due to the orbital motion of the itinerant electrons, it is expressed---through my main assumption---in terms of variables related to the lattice dynamics, bringing in this way the phonons into the description.

An electron rigidly moving with a given ion, as implied by \eqref{xL}---therefore giving the impression of being attached to or localized on that ion---but this electron still being able to wander around as itinerant electrons do, reminds me of the strong correlations intertwining the atomic and band behavior of electrons in transition metals, as first discussed by Hubbard \cite{Hubbard}. These correlations relate to the electron spins, as may be noticed from the following extract from Hubbard's original paper:

``As a guide one may note that Hund's first rule for atoms indicates that the intra-atomic interactions are of such a nature as to align the electron spins on an atom, so one may expect a similar effect in a metal. Suppose now ... that at some instant a given atom has its total spin in the up direction. Then the intra-atomic interactions are, according to Hund's rule, of such a nature that this atom tends to attract electrons with spin up and repel those with spin down. In this way the property of an atom on having total spin at some instant tend to be self-perpetuating ... This persistence of the atomic spin state is not due to the same up-spin electrons being localized on the atom. The actual electrons on the atom are always changing as a result of their band motion, but the electron motions are correlated in such a way as to keep a preponderance of up-spin electrons on the atom. In these circumstances (i.e. if the correlations are strong enough) one can think of the spin as being associated with the atom rather than with the electrons ...''

I believe that the aforementioned phenomenon related to the electron spin should also happen to the orbital degree of freedom of the electrons, with Hund's rule---which is just a consequence of interactions of the form $-\bm{\mu}_{r_e}^{\tm{spin}}\cdot\bm{\mu}_{q_i}^{\tm{spin}}$---replaced by the isotropic part of \eqref{Hoo}, obtained when the first term in \eqref{dk} is neglected. That is, due to interactions of the form $-\bm{\mu}_{r_e}^{\tm{orb}}\cdot\bm{\mu}_{q_i}^{\tm{orb}}$, the energy associated with the isotropic orbital state of the electrons is minimized when a given electron $r_e$ ``belongs'' to an ion $q_i$, the maximum electron-ion attraction occurring when these have the same orbital angular momentum, as assumed in \eqref{xL}---the equality of orbital angular momentum being possible since, as discussed later, the orbital angular momentum of the ions is mass-independent.

In the above statement `` ... minimized when a given electron $r_e$ ``belongs'' to an ion $q_i$'' there is no restriction as to what ion should that electron belong so, under these orbital interactions, it has the freedom to wander from ion to ion as long as the correlations keep a preponderance of conduction electrons sharing the same orbital angular momentum as the ions where they might happen to be at any given moment in time. As, I will show later in this paper, this correlated state of the electrons is particularly possible at temperatures coinciding with that of liquid helium.

In the following I will therefore follow the attitude, that the above conjectured orbital state of the itinerant electrons really takes place in nature, deriving its consequences, as manifested in the electrical resistivity and spin-relaxation. Should this assumption not be true, the current understanding of transport in ferromagnetic metals may be regarded as incomplete, since no quantum-mechanical explanation would then exist for the anomaly discussed in this paper.

\subsection{Mean-field Hamiltonian}

Wrapping up the above discussion, the present model assumes itinerant electrons which are magnetically coupled to lattice ions, from the mentioned approximations to \eqref{Hbi} and \eqref{Hci}, according to
\begin{equation}\label{Hei}
H_{\tm{e-i}}= -\int \bm{\mu}(\bm{x})\cdot [\bm{B}_{i}+\bm{B}_c(\bm{x})]\,d\bm{x},
\end{equation}
where the magnetic moment density due to the itinerant-electron spin is written in second quantization notation as $\bm{\mu}(\bm{x})=-2\mu_B\sum_{ss',\kk\kk'}\varphi_{\kk's'}^{*}(\bm{x})\varphi_{\kk s}(\bm{x})\,\hat{c}_{\kk' s'}^{\dagger}\bm{\sigma}_{s's}\hat{c}_{\kk s}$, with $\varphi_{\kk s}(\bm{x})$ being the Bloch wavefunctions and $\bm{\sigma}_{s's}$ being the Pauli spin-$1/2$ matrices. 

Eq. \eqref{Hei} describes the mean-field approximation to the interaction energy of the conduction-electron spins with the internal magnetic induction, corrected by the spin-orbit interaction arising from these electrons being in ``contact'' with the ions for sufficiently enough time during their band motion. It is now desirable to extract the mean-field effect of the first two terms in \eqref{Hked}.

This leads us to the Stoner model for the itinerant electrons, in which the kinetic and exchange energies of these electrons are described  by \cite{Izuyama2,YosidaB}
\begin{equation}\label{e0}
H_{\tm{kin+ex}} = \sum_{\ks}\varepsilon_{\ks}^0\,n_{\ks},\hspace{0.5cm}\varepsilon_{\bm{k}s}^0=\dfrac{\hbar^2 k^2}{2m}-s\,(\Delta_{\textrm{ex}}/2),
\end{equation}
where $\Delta_{\textrm{ex}}$ is the exchange spin-splitting of the conduction band and, when not a subindex, $s=\pm$ according to a conduction electron having its spin $\uparrow$ or $\downarrow$ with respect to the quantization direction, given by that parallel to the majority spins, i.e., the spin up direction ($-\bm{M}_s/M_s$) for the electrons.

The total Hamiltonian that I consider---taking into account the harmonic displacements of the ions from the lattice positions---is then
\begin{equation*}
 H= \sum_{\ks}\varepsilon_{\ks}^0\,n_{\ks}+\sum_{\qa}\hbar\omega_{\qa} n_{\qa}+H_{\tm{e-i}},
\end{equation*}
where $n_{\qa}=\hat{a}_{\qa}^{\dagger}\hat{a}_{\qa}$ is the operator for the number of phonons with wavevector $\bm{q}$ and polarization $\alpha$. For simplicity, I use the isotropic Debye model, having spectrum with transverse $\omega_{\q 1,2}=\omega_{\q T}=c_T\,q$ and longitudinal $\omega_{\q 3}=\omega_{\q L}=c_L\,q$ excitations, with $c_T$ and $c_L$ the corresponding speeds of sound. 

Extracting the diagonal part of \eqref{Hei} with respect to the Fock basis---due entirely to the Zeeman splitting caused by the internal magnetic induction---the total Hamiltonian can be rewritten as
\begin{equation}\label{Htot}
 H= \sum_{\ks}\varepsilon_{\ks}\,n_{\ks}+\sum_{\qa}\hbar\omega_{\qa} n_{\qa}+H_{\tm{c}},
\end{equation}
where I have denoted $H_c=-\int \bm{\mu}(\bm{x})\cdot \bm{B}_c(\bm{x})d\bm{x}$, and the spin-split conduction-electron bands are
\begin{equation}\label{ef}
\varepsilon_{\bm{k}s}=\varepsilon_{\bm{k}s}^0-s\,(\hbar\omega_c/2),\hspace{0.5cm}\hbar\omega_c=2\mu_B B_{\tm{i}},
\end{equation}
where $\omega_c$ is the conduction-electron cyclotron frequency due to the internal magnetic induction $B_{\tm{i}}=4\pi M_s$ in \eqref{Bi}, with $\mu_B$ being the Bohr magneton. 

The term $H_c$ is nondiagonal with respect to the Fock basis and then causes electron scattering. Furthermore, since both the electronic and the phonon variables appear in this term, it plays the role of an electron-phonon interaction, which is treated here as a perturbation to which the standard perturbation theory in transport phenomena is to be applied.

Before doing this, I anticipate that only the electron scattering events with spin flip contribute to the anomaly sought for---the other processes contributing as $T^3$ to the electrical resistivity---so I only concentrate on these processes. The unit of angular momentum gained or released in such transitions by the itinerant electrons then requires that phonon modes with spin 1 exist, which can give away or absorb that unit of angular momentum. I describe this modes next.   

\subsection{Chiral phonons}
The operator $\bm{\hat{L}}_n=\bm{\hat{r}}_n\times\bm{\hat{p}}_n$ in \eqref{Bc} is the orbital angular momentum (in units of $\hbar$) of the $n^{\tm{th}}$ ion, where $\bm{\hat{r}}_n$ is the displacement of the ion from the lattice point $\bm{R}_n$, and $\bm{\hat{p}}_n=M\bm{\dot{\hat{r}}}_n$, with $M$ the ionic mass. In second quantization notation, this can be expressed \cite{Zhang} as $\bm{\hat{L}}_n=(1/N)\sum_{\bm{qq}'}\bm{\hat{S}}_{\bm{qq}'}e^{i(\bm{q}-\bm{q}')\cdot \bm{R}_n}$, where $N$ is the number of ions and
\begin{equation}\label{Ln}
\begin{split}
 \bm{\hat{S}}_{\bm{qq}'}=\dfrac{1}{2}\sum_{\alpha\alpha'}&\sqrt{\dfrac{\omega_{\qap}}{\omega_{\qa}}}\,(\hat{a}_{\qa}+\hat{a}_{-\qa}^{\dagger})\,\bm{S}_{\q\q'(\alpha\alpha')}\\
&\ \times(\hat{a}_{-\qap}-\hat{a}_{\qap}^{\dagger}),
\end{split}
\end{equation}
with $\hat{a}_{\qa}^{\dagger}$ ($\hat{a}_{\qa}$) being the creation (annihilation) operator of a phonon with wave vector $\q$, angular frequency $\omega_{\qa}$, and \emph{linear} polarization in the direction of the real unit vector $\bm{e}_{\qa}$, and $\bm{S}_{\q\q'(\alpha\alpha')}=-i\,(\hat{\bm{e}}_{\qa}\times\hat{\bm{e}}_{\q'\alpha'})$. Note the independence of \eqref{Ln} on the ionic mass, making plausible referring to $\bm{\hat{L}}_n$ as the orbital angular momentum of a conduction electron rigidly moving with the $n^{\tm{th}}$ ion. 

With the notation  $\bm{S}_{\q(\alpha\alpha')}=\bm{S}_{\q\q(\alpha\alpha')}$, we can write the total angular momentum of the phonon system as $\bm{\hat{S}}_{\textrm{ph}}\equiv\sum_n\bm{\hat{L}}_n=\sum_{\q}\bm{\hat{S}}_{\q,\textrm{ph}}$, where the angular momentum operator of a single phonon, $\bm{\hat{S}}_{\q,\textrm{ph}}=\bm{\hat{S}}_{\q\q}$, with wave vector $\q$ is given by
\begin{equation}
 \bm{\hat{S}}_{\q,\textrm{ph}}=\dfrac{1}{2}\sum_{\alpha\alpha'}\sqrt{\dfrac{\omega_{\qa'}}{\omega_{\qa}}}(\hat{a}_{\qa}+\hat{a}_{-\qa}^{\dagger})\,\bm{S}_{\q(\alpha\alpha')}\,(\hat{a}_{-\qa'}-\hat{a}_{\qa'}^{\dagger}).
\end{equation}
The sum in this expression can be seen as a matrix multiplication. In fact, with $\bm{e}_{\q 1}=(1,0,0)$, $\bm{e}_{\q 2}=(0,1,0)$, and $\bm{e}_{\q 3}=\bm{q}/q=(0,0,1)$, the matrices $\bm{S}_{\q}$ constitute a representation of the infinitesimal generators of rotations in three dimensions, with $\bm{S}_{\q}^{2}=\sum_aS_{\q}^{a}S_{\q}^{a}=2\cdot\bm{1}$, and the commutation relations $\left[ S_{\q}^{a},S_{\q}^{b}\,\right]=i\,\sum_c\epsilon_{abc}S_{\q}^{c}$, and $\left[ S_{\q}^{a},\bm{S}_{\q}^{2}\,\right]=0$. 

As usual in the theory of angular momentum, it is convenient to work in a representation which simultaneously diagonalizes $S_{\q}^{3}$ and $\bm{S}_{\q}^{2}$. This is done by changing basis from $\left\lbrace \bm{e}_{\qa}\right\rbrace$ to the helicity basis $\left\lbrace \bm{\epsilon}_{\qs}\right\rbrace$ defined by the circular $\bm{\epsilon}_{\q\pm}=(1/\sqrt{2})(\bm{e}_{\q 2}\mp i\bm{e}_{\q 1})$ and longitudinal $\bm{\epsilon}_{\q0}=i\bm{e}_{\q 3}$ polarization vectors, where $\bm{\epsilon}_{-\qs}^{*}=\bm{\epsilon}_{\qs}$. The corresponding unitary transformation can be shown to map the matrices $\bm{S}_{\q(\alpha\alpha')}$ to $\bm{\mathcal{S}}_{\q(\sigma\sigma')}=-i\,(\hat{\bm{\epsilon}}_{\qs}^{*}\times\hat{\bm{\epsilon}}_{\qs'})$, which are the spin matrices 

\begin{equation}
\mathcal{S}_{\q}^3 = \begin{pmatrix}
    1 & 0 & 0 \\
    0 & 0 & 0\\
    0 & 0 & -1
\end{pmatrix}
,\hspace{0.1cm}
\mathcal{S}_{\q}^{+} = \begin{pmatrix}
    0 & \sqrt{2} & 0 \\
    0 & 0 & \sqrt{2}\\
    0 & 0 & 0
\end{pmatrix}
,
\hspace{0.1cm} \mathcal{S}_{\q}^{-} = (\mathcal{S}_{\q}^{+})^{\dagger}.
\end{equation}
for a spin-1 particle: the \emph{chiral} phonon. 

In the new representation, the chiral-phonon operators are $\hat{b}_{\q\pm}=(1/\sqrt{2})(\hat{a}_{\q 2}\mp i\hat{a}_{\q 1})$, which annihilate phonons with $\pm$ helicities (circular polarization); and $\hat{b}_{\q0}=i\hat{a}_{\q 3}$, which annihilates zero helicity (or longitudinal) modes. We have, for instance, $\hat{S}_{\q,\textrm{ph}}^3=\hat{b}_{\q+}^{\dagger}\hat{b}_{\q+}-\hat{b}_{\q-}^{\dagger}\hat{b}_{\q-}$. These phonons have definite spin projections $\pm1,0$ along the propagation direction and, making an analogy with circularly polarized light, the displacement field of the ions must rotate perpendicular to the propagation direction in a circularly polarized elastic wave. \cite{Garanin}

It is important to emphasize---and this has long been known\cite{ADlevine} by studying, in the Lagrangian formalism, the transformation properties under rotation of the quantized phonon field---that the spin of the phonon is well-defined in isotropic media and has a value of 1. In a real crystal, it is well-defined only along certain directions of propagation, such as along the lattice vectors of a cubic crystal or along trigonal axes. Since I am using the isotropic Debye model for the phonons in the ferromagnetic metals, phonons with spin 1 are therefore available in all directions for electron scattering.

The magnetic nature of the chiral phonons is not displayed in an isotropic medium in the absence of a external magnetic field---as assumed in the present case---since the transverse bands of the phonon spectrum remain degenerate. For this reason, the itinerant electrons do not experience a further shift of their energy bands due to the contact interaction discussed here, since its diagonal matrix elements in the space of the electrons, being proportional to the thermal-averaged total phonon angular momentum  $\bar{S}_{\textrm{ph}}^3=\bar{N}_{+}-\bar{N}_{-}$ along the quantization direction of the electron spin, \emph{vanish}, due to the same thermal number $\bar{N}_{\pm}$ of right-handed and left-handed phonons caused by the degeneracy.

Therefore, for a direct observation of chiral phonons, the degeneracy of the transverse phonon bands has to be lifted either by the application of an external magnetic field\cite{Holz} or by a spatial symmetry breaking. The latter has been achieved in the recent experiments observing chiral phonons, in a system with broken inversion symmetry of the crystal lattice\cite{HZhu} or with disorder\cite{SKim}. In the present case, however, the circular phonons only play the subsidiary role of being ``reservoirs'' of angular momentum for the electronic spin-flip transitions.

Returning to our technical discussion, I use the convention $\mathcal{PT}\,\hat{b}_{\qs}=\hat{b}_{-\qs}^{*}=\hat{b}_{\qs}$, inherited from $\bm{\epsilon}_{-\qs}^{*}=\bm{\epsilon}_{\qs}$, which states that the wavefunctions of the crystal lattice vibrations are even under the $\mathcal{PT}$ transformation (complex conjugation $+$ parity). When the change to the helicity basis is performed in the tensor product spin space corresponding to $\q$ and $\q'$, I transform $\bm{S}_{\q\q'}\rightarrow\bm{\mathcal{S}}_{\q}\otimes \bm{1}_{\q'}=\bm{1}_{\q}\otimes\bm{\mathcal{S}}_{\q'}$
in \eqref{Ln}, since the image of $\bm{S}_{\q\q'}$ under the transformation must behave as an angular momentum upon rotations and, by definition, must reduce to $\bm{\mathcal{S}}_{\q}$ when $\q=\q'$. 

The ladder operators corresponding to \eqref{Ln}, in the helicity basis, then read
\begin{equation}\label{Sqm}
\begin{split}
  \hat{S}_{\q\q'}^{-}= &\,\dfrac{1}{2}\Biggl[\sqrt{\dfrac{2\omega_{\q'L}}{\omega_{\q T}} }(\hat{b}_{\q-}^{\dagger}+\hat{b}_{\q+})(\hat{b}_{\q'0}^{\dagger}+\hat{b}_{\q'0})\\
  &\ - \sqrt{\dfrac{2\omega_{\q'T}}{\omega_{\q L}} }(\hat{b}_{\q0}^{\dagger}-\hat{b}_{\q0})(\hat{b}_{\q'-}^{\dagger}-\hat{b}_{\q'+})\Biggr],
\end{split}
\end{equation}
and $\hat{S}_{\q\q'}^{+}=(\hat{S}_{\q\q'}^{-})^{\dagger}$ which enter the terms causing electron spin-flip scattering.

By using the Bloch theorem for $\varphi_{\kk s}(\bm{x})=e^{i\kk \cdot \bm{x}}u_{\kk s}(\bm{x})$, that is, $u_{\kk s}(\bm{x+\bm{R}_n})=u_{\kk s}(\bm{x})$ to express $u_{\kk s}(\bm{\bm{R}_n})=u_{\kk s}(0)=\varphi_{\kk s}(0)$, and neglecting umklapp processes, the relevant terms in $H_c$ are mainly
\begin{equation}\label{Hc}
 \frac{1}{2}\sum_{\kk \kk', \q \q'} A_{\kk \downarrow,\kk'\uparrow}\,\delta_{\Delta\q,-\Delta \kk}\;\hat{c}_{\kk'\uparrow}^{\dagger}\hat{c}_{\kk \downarrow}\;\hat{S}_{\bm{qq}'}^{-}, 
\end{equation}
where $\Delta \kk=\kk'-\kk$, $\Delta\q=\q'-\q$, and the matrix elements $A_{\kk s,\kk' s'}=-\textstyle{\frac{16\pi}{3}}\mu_B^2 \varphi_{\kk' s'}^{*}(0)\varphi_{\kk s}(0)$ give the strength of the resulting electron-phonon interaction. These terms account for processes where spin-$\downarrow$ electrons transition to the majority-spin conduction band. The reverse processes can be shown to be exponentially suppressed at low temperatures.

\section{Results and discussion}
The transition probability rate for a conduction electron undergoing a collision from the state $\kk\hspace{-0.1cm}\downarrow$ to an unoccupied state $\kk'\hspace{-0.1cm}\uparrow$ is obtained from \eqref{Sqm} and \eqref{Hc}, in the leading order of perturbation theory, using Fermi's golden rule. \cite{Solano2} Since for temperatures $T\rightarrow0$ the average occupation of a phonon mode $\bar{N}_{\q\sigma}$ is exponentially small \emph{except} for the lowest energy mode supported by the crystal lattice---with wavevector magnitude $q_{\tm{min}}=2\pi/L=0^{+}$, where $L$ is the largest linear size of the sample---the processes happening more frequently at very low temperatures are those in which these lowest energy modes are involved, for which conservation of energy reads
\begin{equation}\label{cE}
\hbar c_T |\Delta\kk| -|\Delta \varepsilon_{\kk}|=0,
\end{equation}
up to a negligible term of $O(1/L)$. Therefore, neglecting contributions from exponentially small terms as well as the temperature-independent term (discussed later), the transition probability rate per electron, for temperatures $T\rightarrow0$, is 
\begin{equation}\label{w}
\begin{split}
 w_{\kk\downarrow\rightarrow\kk'\uparrow}&=\dfrac{3\pi}{16}\dfrac{|A_{\kk \downarrow,\kk' \uparrow}|^2|\Delta\epsilon_{\kk}|^2}{(k_B\Theta)^4}\dfrac{k_F}{|\Delta \kk|}\dfrac{c_s^4}{c_L^2c_T^2}\left(\dfrac{2}{z}\right)^{4/3}\\
&\  \times \dfrac{(k_BT)}{\hbar}\ln\left(\dfrac{k_BT}{\hbar c_L q_{\tm{min}}}\right)+O(T^2),
 \end{split}
\end{equation}
where $\Theta$ is the Debye temperature, $k_F$ is the Fermi wave vector, $z=N_e/N$ is the number of conduction electrons per ion and $c_s$ is average speed of sound, determined from $3/c_s^3=2/c_T^3+1/c_L^3$.

The dominant (first) term in \eqref{w} comes from processes wherein a lowest-energy longitudinal mode of the lattice is absorbed and a circular mode is \emph{spontaneously} emitted to satisfy the conservation laws. The most probable transitions per unit time are obtained from \eqref{w} when $|\Delta\epsilon_{\kk}|$ is maximum and $|\Delta \kk|$ is minimum, corresponding respectively to $|\Delta\epsilon_{\kk}|=|E_{F\uparrow}-E_{F\downarrow}|=\hbar\omega_c$ and $|\Delta \kk|=k_{F\uparrow}-k_{F\downarrow}\equiv\Delta k_F$.

The most probable processes then satisfy \eqref{cE} in the form $\hbar \omega_c=\hbar c_T \Delta k_F$ or, equivalently, by defining the characteristic temperature $T_{\tm{i}}=\hbar \omega_c/k_B$ associated with the internal magnetic induction and $T_{\tm{res}}=\hbar c_T \Delta k_F/k_B$ as that associated with the aforementioned spontaneous excitation of circular-phonon modes, \eqref{cE} may be written in the form $T_{\tm{i}}=T_{\tm{res}}$. By writing \cite{Yosida} $\Delta  k_{F}/k_F=\Delta_{\tm{ex}}/2E_F$, we can express $T_{\tm{res}}$ as
\begin{equation}\label{Ts}
 T_{\tm{res}} = \dfrac{c_T}{c_s}(z/2)^{1/3}(\Delta_{\tm{ex}}/2E_F)\,\Theta,
\end{equation}
whose agreement with $T_{\tm{i}}$, as shown in table \ref{t1}, is remarkably good for the pure ferromagnetic metals, with the characteristic temperatures involved being around liquid-helium temperatures. 
\begin{table*}[t]
\begin{center}
    \begin{tabular}{cccccccccccccc}
    \ \vspace{-3mm}\\\rowcolor{snow}
    Metal & $z$ & $n$ $(10^{28}$m$^{-3})$ & $c_L$ $(m/s)$ & $T_{\tm{res}}$ ($^{\circ}$K) & $T_{\tm{i}}$ ($^{\circ}$K) & $\Theta$ ($^{\circ}$K) & $H_a$(MG) &$\xi$ & $\delta\Omega_F$ & \multicolumn{2}{c}{$\gamma$ ($10^{-12}\Omega$ cm/$^{\circ}$K)}& \multicolumn{2}{c}{$\tau_{\downarrow\uparrow}$ (ns)} \\ \rowcolor{snow}
     & & & $c_T=c_L/\sqrt{2}$& & & & & & & theory & exp. & theory & exp. \\\rowcolor{lgray}
     Fe&  0.22 & 8.5 & 5960& 4.0 & 2.9 &470 & 1.95 &0.35& 0.5&16.8&11$-$49.3& 0.04&0.03$-$0.1\\ \rowcolor{gwhite}
     Co & 0.7 & 9.1 & 4720& 5.4 & 2.4 &445 & 2.2&0.21&8.0 & 6.4&3$-$32& 0.03&$-$\\\rowcolor{lgray}           
     Ni & 0.6& 9.2 &6040& 1.0 & 0.8 &450 & 2.4 &0.55& 8.0&7.9&5.8$-$16& 0.03&0.02$-$0.1       
\\
    \end{tabular}
\caption{Predicted against observed values of the linear coefficient $\gamma$ (from \eqref{gamma}) in the electrical resistivity of the ferromagnetic metals at low temperatures, and the spin-lattice relaxation time $\tau_{\downarrow\uparrow}$ at $T=300^{\circ}$K (from \eqref{utau}). See the text for a discussion of the appropriate values of the other quantities presented here.  \label{t1}}
\end{center}
\end{table*}

In order to estimate the spin-lattice relaxation time corresponding to the transition rates in \eqref{w}, it is necessary to account for less probable transitions involving conduction-electron energy losses and crystal momentum transfers in the ranges $0\le|\Delta\epsilon_{\kk}|<\hbar\omega_c$ and $\Delta k_F<|\Delta \kk|\le 2k_F $, respectively. Since only order of magnitudes are of interest here, no more sophistication than averaging \eqref{w} over the solid angle is required. That is $1/\tau_{\downarrow\uparrow}\equiv4\pi \delta\Omega_F\,\tm{mean}(\avg{w_{\kk\downarrow\rightarrow\kk'\uparrow}^{\tm{max}}},\avg{w_{\kk\downarrow\rightarrow\kk'\uparrow}^{\tm{min}}})=2\pi\delta\Omega_F\,\avg{w_{\kk_F\downarrow\rightarrow\kk_F\uparrow}}$, where the average $\avg{\cdot}$ is taken over the solid angle between $\kk$ and $\kk'$.

The factor $\delta\Omega_F$ is included to account for the realistic reduction ($\delta\Omega_F<1$) or increase ($\delta\Omega_F>1$) in effective solid angle from that subtended by a spherical Fermi surface ($\delta\Omega_F=1$). Since $\avg{1/|\Delta \kk|}=1/\tm{max}(k,k')$ in \eqref{w}, the factor $\delta\Omega_F$ is understood as referring to the majority-spin Fermi surface shape. With this, the spin-lattice relaxation time is
\begin{equation}\label{utau}
\begin{split}
 \dfrac{1}{\tau_{\downarrow\uparrow}}&=\dfrac{3\pi^2}{8}\delta\Omega_F\dfrac{\avg{|A_{k_F\downarrow,k_F\uparrow}|^2}(\hbar\omega_c)^2}{(k_B\Theta)^4}\left(\dfrac{2}{z}\right)^{4/3}\dfrac{c_s^4}{c_L^2c_T^2}\\
 &\ \times \dfrac{(k_BT)}{\hbar}\ln\left(\dfrac{k_BT}{\hbar c_L q_{\tm{min}}}\right)+O(T^2).
 \end{split}
\end{equation}

Due to the isotropy of the model, the contribution of the discussed scattering mechanism to the ideal electrical resistivity (excluding the residual term at $T=0$, discussed in the Appendix) of the ferromagnetic metals at low temperatures can be calculated from the Drude-Mott formula \cite{Mott3}
\begin{equation}\label{rhop}
  \rho(T)=\dfrac{m}{e^2z\, n}\dfrac{1}{\tau_{\uparrow\downarrow}}=\gamma\,T + \eta\left(\dfrac{T}{\Theta}\right)\ln\left(\dfrac{T}{\Theta}\right)+O(T^2),
\end{equation}
with $n=N/V$ the atomic density and the coefficients $\gamma$ and $\eta$ readily obtained from \eqref{utau}, the first term being the dominant term around liquid-helium temperatures since $T/\Theta$ is then much smaller than $c_sq_D/c_Lq_{\tm{min}}$, with $q_D$ being he Debye cutoff---I use a typical length $L= 35$ mm for the largest dimension of the samples in the electrical resistivity measurements. \cite{Semenenko} 

\subsection{Numerical estimates}
In order to obtain numerical estimates, let me discuss the values of the physical quantities involved in \eqref{utau} and \eqref{rhop}, which are described, in the following, in the order shown in Table \ref{t1} (from left to right). The number of conduction electrons per ion $z$ consistent with the observed values of the saturation magnetization, are taken from ref. [\onlinecite{Mott}]; the atomic densities $n$ as well as the Fermi energies $E_F\simeq 7$ eV for the three elements are taken from ref. [\onlinecite{Goodings}]; the longitudinal speeds of sound $c_L$ are taken from ref. [\onlinecite{Lide}], and I use $c_T=c_L/\sqrt{2}$, which is valid for elastically isotropic bodies (coming from the corresponding relation \cite{Fuchs} $c_{11}=2\,c_{44}$ for the elastic constants, with $c_{12}=0$), for consistency with the isotropic model. 

The exchange splittings of the $4s$ band $\Delta_{\tm{ex}}$ are estimated as the energy difference between the bottoms of the majority-spin and minority-spin bands at the $\Gamma$ point (center of Brillouin zone) extracted from the band-structure calculations of ref. [\onlinecite{Nautiyal}] for Fe, ref. [\onlinecite{Singal}] for Co and ref. [\onlinecite{Wang}] for Ni,  with the approximate values 0.13, 0.27 and 0.05 eV, respectively; from these the temperatures $T_{\tm{res}}$ are obtained from Eq. \eqref{Ts}. The temperatures $T_{\tm{i}}$ have been estimated from the well-known\cite{Berger} magnitudes of the internal magnetic induction $B_{\textrm{i}}$ of 22, 18, and 6 kG for Fe, Co and Ni, respectively. The low temperature limits of the Debye temperature $\Theta$ are taken from ref. [\onlinecite{Kittel}], and the outermost $s$-electron hyperfine fields in the free atom $H_a$ from ref. [\onlinecite{Watson}].

It is convenient to write $\avg{|A_{k_F\uparrow,k_F\downarrow}|}=2\mu_B \xi H_a$, with the Knight ratio \cite{Townes} defined as $\xi=\langle|\bar{\varphi}_{k_F}(0)|^2\rangle/|\psi_a(0)|^2$, where $\langle|\bar{\varphi}_{k_F}(0)|^2\rangle$ is the average probability density at the nucleus of electronic states on the Fermi surface, and $\psi_a(0)$ the wavefunction at the nucleus of the outermost $s$ electron in the free atom which, as it is known, produces a hyperfine field of magnitude $H_a=(8\pi/3)\mu_B|\psi_a(0)|^2$. Note that, in writing $\avg{|A_{k_F\uparrow,k_F\downarrow}|}$ in this form, I have assumed that $\varphi_{k_F\uparrow}(0)$ and $\varphi_{k_F\downarrow}(0)$ deviate only slightly \cite{Song,Duff2} from their arithmetic mean $\bar{\varphi}_{k_F}(0)$.

The Knight ratio accounts for any deviation in hyperfine coupling from atomic behavior and may deviate from $\xi=1$ for two reasons: \cite{Bennett2} (i) the reduction of $s$-character of the wavefunctions at the Fermi surface and (ii) the fact that the wavefunctions in a metal are normalized within volumes smaller than in the free atom, causing the conduction electron density in the metal greater than in the free atom. For ``simple'' metals the reduction tends to predominate over the normalization effect, with $\xi$ taking values between 0.1 and 0.8. 

We take both effects into account, in their simplest form, by taking $\xi=(1/N_{3d4s})a^3/(4\pi r^3/3)$, with $N_{3d4s}$ the number of electrons per atom in the $3d$ and $4s$ subshells, with values 8, 9, and 10 for Fe, Co, and Ni, respectively, $a$ is the lattice constant with values 2.87, 2.51 and 3.52 \AA{} for Fe, Co and Ni, respectively \cite{Kittel}, and $r$ is the atomic radius with values 1.26, 1.25 and 1.24 \AA{} for Fe, Co and Ni, respectively \cite{Kittel}.  

Since $|\bar{\varphi}|^2$ and $|\psi|^2$ have units of 1/volume, the normalization effect is then taken into account by the ratio of unit cell to atomic volumes $a^3/(4\pi r^3/3)$, and the reduction in $s$-character by the factor $1/N_{3d4s}$, the latter because of the $N_{3d4s}$ electrons in the $3d$ and $4s$ subshells of the free atom, only a fraction $z$ remains in the $4s$ band in the solid, \cite{Mott} and we need to count all the electrons which can make this donation to the $4s$ band. 

With this simple rule, the values of $\xi$ shown in Table \ref{t1} are in the range of those for ``simple'' metals. Moreover, the $4s$ contribution to the effective hyperfine fields $\xi H_a$ obtained in this way are in agreement (except for Ni, for which it is an order of magnitude higher) with the theoretical estimates of ref. [\onlinecite{Muto}] aimed at explaining the observed hyperfine fields in the $3d$ ferromagnetic metals from M\"{o}ssbauer and NMR experiments. 

Finally, we need to estimate the factors $\delta \Omega_F$. For Ni (fcc structure), the sheet of the Fermi surface coming from the $4s$ majority-spin band is, as in copper, spherical-like,  with necks touching the Brillouin zone faces near the $L$ points \cite{Gold}. A \emph{rough} picture of this surface may then be drawn by considering the union of a major sphere of radius $k_{F\uparrow}$, centered at the $\Gamma$ point, with little spheres with such radii as to touch the major sphere and the $L$ points, i.e. the centers of the hexagonal faces of the truncated octahedron constituting the Brillouin zone of a fcc structure. Since there are 8 such faces, we have 8 little spheres and then we need to multiply the value $\delta\Omega_F=1$ corresponding to the major sphere by 8, as shown in Table \ref{t1}. 

For Co (hcp structure), similar results apply due to the correspondence between energy bands (and Fermi surfaces) of the fcc and hcp structures when the hcp double zone is rotated until its [0001] axis coincides with the [111] axis of the fcc zone \cite{Gold}. For Fe (bcc structure), the situation is much more complex since the Fermi surface corresponding to the $4s$-band breaks up into small regions of electron and hole pockets \cite{Gold,Stearns}. Nevertheless, I consider this as a reduction in effective solid angle and take for $\delta\Omega_F$ the neutral value shown in Table \ref{t1}, having in mind that a more detailed investigation should not considerably change the overall result. 

\subsection{Comparison with experiments}
Having described the magnitudes of the relevant quantities defining the linear coefficient
\begin{equation}\label{gamma}
\begin{split}
 \gamma&=\dfrac{3\pi^2}{4}\delta\Omega_F\dfrac{(\mu_B \xi H_a)^2(\hbar\omega_c)^2}{(k_B\Theta)^4}\left(\dfrac{2}{z}\right)^{7/3}\dfrac{c_s^4}{c_L^2c_T^2}\\
 &\ \hspace{1cm}\times \dfrac{mk_B}{e^2n\hbar}\ln\left(\dfrac{k_B\Theta}{\hbar c_L q_{\tm{min}}}\right),
 \end{split}
\end{equation}
in \eqref{rhop}, I show the theoretical estimates from the present model in Table \ref{t1}, as well as the predicted spin-lattice relaxation time $\tau_{\downarrow\uparrow}$ at $T=300^{\circ}$K from \eqref{utau}. These are compared with the experimental values. 

The observed values of the coefficient $\gamma$ from electrical resistivity measurements are taken from the review article of\; \textcite{Volkenshtein}. The ranges shown represent the minimum and maximum values of measurements of the linear coefficient by multiple authors under similar experimental conditions. In the three materials, the predicted values then agree with the measurements within the statistical error.

As for the spin-lattice relaxation time at 300$^{\circ}$K, from the line width of the ferromagnetic resonance signal a value of about $0.1$ ns is found to fit the data best \cite{Bloembergen2,Holmes} for Fe and Ni, although magneto-optic Kerr effect measurements \cite{Agranat} reveal a spin-lattice relaxation time between 20 ps and 40 ns for Ni, and spin-polarized photoemission experiments \cite{Vaterlaus} reveal this to be greater than 30 ps for Fe. These measurements are then shown as ranges in Table \ref{t1}, within which the theoretical predictions fall again. I was not able to find measurements of the spin-lattice relaxation time of hcp cobalt but, from the theoretical estimates, this is believed to be nearly the same as that of iron and nickel. 

The agreement of the theoretical predictions with the observed values from independent experiments and for different materials then indicate that the main assumption of this paper about the correlated motion of the itinerant electrons is most likely to be true. This, however, should be subjected to further studies.

\begin{acknowledgments}
I would like to acknowledge the support from the Fulbright-Colciencias Fellowship, the Columbia GSAS Faculty Fellowship, and express my gratitude to Prof. Andrew Millis for valuable discussions. 
\end{acknowledgments}

\appendix

\section{Temperature-independent contribution to the electrical resistivity}
The temperature-independent contributions to the spin-lattice relaxation time and electrical resistivity come from processes where the pair of chiral phonons involved in each electron spin-flip scattering event are emitted spontaneously. Defining the corresponding contribution to the spin-lattice relaxation time as in the temperature-dependent case, that is, by the average $1/\tau_{\downarrow\uparrow}^{0}=4\pi\delta\Omega_F\,\tm{mean}(\avg{w_{\kk\downarrow\rightarrow\kk'\uparrow}^{\tm{sp-sp (max)}}},\avg{w_{\kk\downarrow\rightarrow\kk'\uparrow}^{\tm{sp-sp (min)}}})$, it is easily shown that
\begin{equation}\label{ut0}
 \dfrac{1}{\tau_{\downarrow\uparrow}^{0}}=\dfrac{3\pi^2}{8}\,\delta\Omega_F\,\dfrac{\langle|A_{k_F \downarrow,k_F\uparrow}|^2\rangle(\hbar\omega_c)^3}{\hbar(k_B\Theta)^4}\dfrac{c_s^4}{c_T^2c_L^2}\left(\dfrac{2}{z}\right)^{4/3}.
\end{equation}
This is of the order of $10^{-7}$s for Fe, Co and Ni and then the total leading contribution $\tau_{\downarrow\uparrow}^{\tm{tot}}$ to the spin-lattice relaxation time, defined as $1/\tau_{\downarrow\uparrow}^{\tm{tot}}=1/\tau_{\downarrow\uparrow}+1/\tau_{\downarrow\uparrow}^{0}$, is dominated around liquid-helium temperatures by the temperature-dependent part, as shown in the main article, since $\tau_{\downarrow\uparrow}^{0}/\tau_{\downarrow\uparrow}\sim 10^{4}$. 

It is well-known that, for the ferromagnetic metals, the residual resistivity is not only due to impurities and other lattice defects but that an anomalous contribution from the internal magnetic induction is also present, \cite{Volkenshtein,Berger} apart from magnetostriction and magnetocrystal contributions. As far as I know, there has been no theory so far describing that contribution from the internal magnetic induction. From \eqref{ut0} this is given by
\begin{equation}
 \rho_0= \dfrac{m}{e^2zn}\dfrac{1}{\tau_{\downarrow\uparrow}^{0}},
\end{equation}
which is of the order of $10^{-12}\,\Omega$ cm for Fe, Co and Ni.

\bibliography{references}

\begin{thebibliography}{53}%
\makeatletter
\providecommand \@ifxundefined [1]{%
 \@ifx{#1\undefined}
}%
\providecommand \@ifnum [1]{%
 \ifnum #1\expandafter \@firstoftwo
 \else \expandafter \@secondoftwo
 \fi
}%
\providecommand \@ifx [1]{%
 \ifx #1\expandafter \@firstoftwo
 \else \expandafter \@secondoftwo
 \fi
}%
\providecommand \natexlab [1]{#1}%
\providecommand \enquote  [1]{``#1''}%
\providecommand \bibnamefont  [1]{#1}%
\providecommand \bibfnamefont [1]{#1}%
\providecommand \citenamefont [1]{#1}%
\providecommand \href@noop [0]{\@secondoftwo}%
\providecommand \href [0]{\begingroup \@sanitize@url \@href}%
\providecommand \@href[1]{\@@startlink{#1}\@@href}%
\providecommand \@@href[1]{\endgroup#1\@@endlink}%
\providecommand \@sanitize@url [0]{\catcode `\\12\catcode `\$12\catcode
  `\&12\catcode `\#12\catcode `\^12\catcode `\_12\catcode `\%12\relax}%
\providecommand \@@startlink[1]{}%
\providecommand \@@endlink[0]{}%
\providecommand \url  [0]{\begingroup\@sanitize@url \@url }%
\providecommand \@url [1]{\endgroup\@href {#1}{\urlprefix }}%
\providecommand \urlprefix  [0]{URL }%
\providecommand \Eprint [0]{\href }%
\providecommand \doibase [0]{http://dx.doi.org/}%
\providecommand \selectlanguage [0]{\@gobble}%
\providecommand \bibinfo  [0]{\@secondoftwo}%
\providecommand \bibfield  [0]{\@secondoftwo}%
\providecommand \translation [1]{[#1]}%
\providecommand \BibitemOpen [0]{}%
\providecommand \bibitemStop [0]{}%
\providecommand \bibitemNoStop [0]{.\EOS\space}%
\providecommand \EOS [0]{\spacefactor3000\relax}%
\providecommand \BibitemShut  [1]{\csname bibitem#1\endcsname}%
\let\auto@bib@innerbib\@empty
\bibitem [{\citenamefont {Campbell}\ and\ \citenamefont
  {Fert}(1982)}]{Campbell}%
  \BibitemOpen
  \bibfield  {author} {\bibinfo {author} {\bibfnamefont {I.~A.}\ \bibnamefont
  {Campbell}}\ and\ \bibinfo {author} {\bibfnamefont {A.}~\bibnamefont
  {Fert}},\ }\href@noop {} {\emph {\bibinfo {title} {{Ferromagnetic Materials,
  E. P. Wohlfarth}}}},\ Vol.~\bibinfo {volume} {3}\ (\bibinfo  {publisher}
  {North-Holland},\ \bibinfo {address} {Amsterdam},\ \bibinfo {year}
  {1982})\BibitemShut {NoStop}%
\bibitem [{\citenamefont {Volkenshtein}\ \emph {et~al.}(1973)\citenamefont
  {Volkenshtein}, \citenamefont {Dyakina},\ and\ \citenamefont
  {Startsev}}]{Volkenshtein}%
  \BibitemOpen
  \bibfield  {author} {\bibinfo {author} {\bibfnamefont {N.~V.}\ \bibnamefont
  {Volkenshtein}}, \bibinfo {author} {\bibfnamefont {V.~P.}\ \bibnamefont
  {Dyakina}}, \ and\ \bibinfo {author} {\bibfnamefont {V.~E.}\ \bibnamefont
  {Startsev}},\ }\bibfield  {title} {\enquote {\bibinfo {title} {{Scattering
  Mechanisms of Conduction Electrons in Transition Metals at Low
  Temperatures}},}\ }\href@noop {} {\bibfield  {journal} {\bibinfo  {journal}
  {phys. stat. sol. (b)}\ }\textbf {\bibinfo {volume} {57}},\ \bibinfo {pages}
  {9} (\bibinfo {year} {1973})}\BibitemShut {NoStop}%
\bibitem [{\citenamefont {Kasuya}(1959)}]{Kasuya2}%
  \BibitemOpen
  \bibfield  {author} {\bibinfo {author} {\bibfnamefont {T.}~\bibnamefont
  {Kasuya}},\ }\bibfield  {title} {\enquote {\bibinfo {title} {{Effects of
  $s$-$d$ Interaction on Transport Phenomena}},}\ }\href@noop {} {\bibfield
  {journal} {\bibinfo  {journal} {Prog. Theor. Phys.}\ }\textbf {\bibinfo
  {volume} {22}},\ \bibinfo {pages} {227} (\bibinfo {year} {1959})}\BibitemShut
  {NoStop}%
\bibitem [{\citenamefont {Goodings}(1963)}]{Goodings}%
  \BibitemOpen
  \bibfield  {author} {\bibinfo {author} {\bibfnamefont {D.~A.}\ \bibnamefont
  {Goodings}},\ }\bibfield  {title} {\enquote {\bibinfo {title} {{Electrical
  Resistivity of Ferromagnetic Metals at Low Temperatures}},}\ }\href@noop {}
  {\bibfield  {journal} {\bibinfo  {journal} {Phys. Rev.}\ }\textbf {\bibinfo
  {volume} {132}},\ \bibinfo {pages} {542} (\bibinfo {year}
  {1963})}\BibitemShut {NoStop}%
\bibitem [{\citenamefont {Mannari}(1959)}]{Mannari}%
  \BibitemOpen
  \bibfield  {author} {\bibinfo {author} {\bibfnamefont {I.}~\bibnamefont
  {Mannari}},\ }\bibfield  {title} {\enquote {\bibinfo {title} {{Electrical
  Resistance of Ferromagnetic Metals}},}\ }\href@noop {} {\bibfield  {journal}
  {\bibinfo  {journal} {Prog. Theor. Phys.}\ }\textbf {\bibinfo {volume}
  {22}},\ \bibinfo {pages} {335} (\bibinfo {year} {1959})}\BibitemShut
  {NoStop}%
\bibitem [{\citenamefont {Baber}(1937)}]{Baber}%
  \BibitemOpen
  \bibfield  {author} {\bibinfo {author} {\bibfnamefont {W.~G.}\ \bibnamefont
  {Baber}},\ }\bibfield  {title} {\enquote {\bibinfo {title} {{The Contribution
  to the Electrical Resistance of Metals from Collisions between Electrons}},}\
  }\href@noop {} {\bibfield  {journal} {\bibinfo  {journal} {Proc. Roy. Soc.
  (London)}\ }\textbf {\bibinfo {volume} {A158}},\ \bibinfo {pages} {383}
  (\bibinfo {year} {1937})}\BibitemShut {NoStop}%
\bibitem [{\citenamefont {Turov}(1955{\natexlab{a}})}]{Turov}%
  \BibitemOpen
  \bibfield  {author} {\bibinfo {author} {\bibfnamefont {E.~A.}\ \bibnamefont
  {Turov}},\ }\bibfield  {title} {\enquote {\bibinfo {title} {{Relaxation
  processes in ferromagnetic metals at low temperatures}},}\ }\href@noop {}
  {\bibfield  {journal} {\bibinfo  {journal} {Izv. Akad. Nauk. SSSR, Ser.
  Fiz.}\ }\textbf {\bibinfo {volume} {19}},\ \bibinfo {pages} {462} (\bibinfo
  {year} {1955}{\natexlab{a}})}\BibitemShut {NoStop}%
\bibitem [{\citenamefont {Turov}(1955{\natexlab{b}})}]{Turov2}%
  \BibitemOpen
  \bibfield  {author} {\bibinfo {author} {\bibfnamefont {E.~A.}\ \bibnamefont
  {Turov}},\ }\bibfield  {title} {\enquote {\bibinfo {title} {{Electrical
  conductivity of ferromagnetic metals at low temperatures}},}\ }\href@noop {}
  {\bibfield  {journal} {\bibinfo  {journal} {Izv. Akad. Nauk. SSSR, Ser.
  Fiz.}\ }\textbf {\bibinfo {volume} {19}},\ \bibinfo {pages} {474} (\bibinfo
  {year} {1955}{\natexlab{b}})}\BibitemShut {NoStop}%
\bibitem [{\citenamefont {Abelskii}\ and\ \citenamefont
  {Turov}(1960)}]{Turov3}%
  \BibitemOpen
  \bibfield  {author} {\bibinfo {author} {\bibfnamefont {Sh.~Sh.}\ \bibnamefont
  {Abelskii}}\ and\ \bibinfo {author} {\bibfnamefont {E.~A.}\ \bibnamefont
  {Turov}},\ }\bibfield  {title} {\enquote {\bibinfo {title} {{To the theory of
  temperature dependence of electrical conductivity and thermal conductivity of
  ferromagnets at low temperatures}},}\ }\href@noop {} {\bibfield  {journal}
  {\bibinfo  {journal} {Fiz. Metal. Metalloved}\ }\textbf {\bibinfo {volume}
  {10}},\ \bibinfo {pages} {801} (\bibinfo {year} {1960})}\BibitemShut
  {NoStop}%
\bibitem [{\citenamefont {Taylor}\ \emph {et~al.}(1968)\citenamefont {Taylor},
  \citenamefont {Isin},\ and\ \citenamefont {Coleman}}]{Taylor}%
  \BibitemOpen
  \bibfield  {author} {\bibinfo {author} {\bibfnamefont {G.~R.}\ \bibnamefont
  {Taylor}}, \bibinfo {author} {\bibfnamefont {A.}~\bibnamefont {Isin}}, \ and\
  \bibinfo {author} {\bibfnamefont {R.~V.}\ \bibnamefont {Coleman}},\
  }\bibfield  {title} {\enquote {\bibinfo {title} {{Resistivity of Iron as a
  Function of Temperature and Magnetization}},}\ }\href@noop {} {\bibfield
  {journal} {\bibinfo  {journal} {Phys. Rev.}\ }\textbf {\bibinfo {volume}
  {165}},\ \bibinfo {pages} {621} (\bibinfo {year} {1968})}\BibitemShut
  {NoStop}%
\bibitem [{\citenamefont {Overhauser}(1953)}]{Overhauser}%
  \BibitemOpen
  \bibfield  {author} {\bibinfo {author} {\bibfnamefont {A.~W.}\ \bibnamefont
  {Overhauser}},\ }\bibfield  {title} {\enquote {\bibinfo {title}
  {{Paramagnetic Relaxation in Metals}},}\ }\href@noop {} {\bibfield  {journal}
  {\bibinfo  {journal} {Phys. Rev.}\ }\textbf {\bibinfo {volume} {89}},\
  \bibinfo {pages} {689} (\bibinfo {year} {1953})}\BibitemShut {NoStop}%
\bibitem [{\citenamefont {Fabian}\ and\ \citenamefont {{Das
  Sarma}}(1999)}]{Fabian}%
  \BibitemOpen
  \bibfield  {author} {\bibinfo {author} {\bibfnamefont {J.}~\bibnamefont
  {Fabian}}\ and\ \bibinfo {author} {\bibfnamefont {S.}~\bibnamefont {{Das
  Sarma}}},\ }\bibfield  {title} {\enquote {\bibinfo {title} {{Spin relaxation
  of conduction electrons}},}\ }\href@noop {} {\bibfield  {journal} {\bibinfo
  {journal} {J. Vac. Sci. Technol. B}\ }\textbf {\bibinfo {volume} {17}},\
  \bibinfo {pages} {1708} (\bibinfo {year} {1999})}\BibitemShut {NoStop}%
\bibitem [{\citenamefont {Boross}\ \emph {et~al.}(2013)\citenamefont {Boross},
  \citenamefont {D{\'o}ra}, \citenamefont {Kiss},\ and\ \citenamefont
  {Simon}}]{Boross}%
  \BibitemOpen
  \bibfield  {author} {\bibinfo {author} {\bibfnamefont {P.}~\bibnamefont
  {Boross}}, \bibinfo {author} {\bibfnamefont {B.}~\bibnamefont {D{\'o}ra}},
  \bibinfo {author} {\bibfnamefont {A.}~\bibnamefont {Kiss}}, \ and\ \bibinfo
  {author} {\bibfnamefont {F.}~\bibnamefont {Simon}},\ }\bibfield  {title}
  {\enquote {\bibinfo {title} {{A unified theory of spin-relaxation due to
  spin-orbit coupling in metals and semiconductors}},}\ }\href@noop {}
  {\bibfield  {journal} {\bibinfo  {journal} {Sci. Rep.}\ }\textbf {\bibinfo
  {volume} {3}},\ \bibinfo {pages} {3233} (\bibinfo {year} {2013})}\BibitemShut
  {NoStop}%
\bibitem [{\citenamefont {Mokrousov}\ \emph {et~al.}(2013)\citenamefont
  {Mokrousov}, \citenamefont {Zhang}, \citenamefont {Freimuth}, \citenamefont
  {Zimmermann}, \citenamefont {Long}, \citenamefont {Weischenberg},
  \citenamefont {Souza}, \citenamefont {Mavropoulos},\ and\ \citenamefont
  {Bl{\"u}gel}}]{Mokrousov}%
  \BibitemOpen
  \bibfield  {author} {\bibinfo {author} {\bibfnamefont {Y.}~\bibnamefont
  {Mokrousov}}, \bibinfo {author} {\bibfnamefont {H.}~\bibnamefont {Zhang}},
  \bibinfo {author} {\bibfnamefont {F.}~\bibnamefont {Freimuth}}, \bibinfo
  {author} {\bibfnamefont {B.}~\bibnamefont {Zimmermann}}, \bibinfo {author}
  {\bibfnamefont {N.~H.}\ \bibnamefont {Long}}, \bibinfo {author}
  {\bibfnamefont {J.}~\bibnamefont {Weischenberg}}, \bibinfo {author}
  {\bibfnamefont {I.}~\bibnamefont {Souza}}, \bibinfo {author} {\bibfnamefont
  {P.}~\bibnamefont {Mavropoulos}}, \ and\ \bibinfo {author} {\bibfnamefont
  {S.}~\bibnamefont {Bl{\"u}gel}},\ }\href@noop {} {\bibfield  {journal}
  {\bibinfo  {journal} {J. Phys: Condens. Matter}\ }\textbf {\bibinfo {volume}
  {25}},\ \bibinfo {pages} {163201} (\bibinfo {year} {2013})}\BibitemShut
  {NoStop}%
\bibitem [{\citenamefont {Zhang}\ and\ \citenamefont {Niu}(2014)}]{Zhang}%
  \BibitemOpen
  \bibfield  {author} {\bibinfo {author} {\bibfnamefont {L.}~\bibnamefont
  {Zhang}}\ and\ \bibinfo {author} {\bibfnamefont {Q.}~\bibnamefont {Niu}},\
  }\bibfield  {title} {\enquote {\bibinfo {title} {{Angular Momentum of Phonons
  and the Einstein-de Haas Effect}},}\ }\href@noop {} {\bibfield  {journal}
  {\bibinfo  {journal} {Phys. Rev. Lett.}\ }\textbf {\bibinfo {volume} {112}},\
  \bibinfo {pages} {085503} (\bibinfo {year} {2014})}\BibitemShut {NoStop}%
\bibitem [{\citenamefont {Zhang}\ and\ \citenamefont {Niu}(2015)}]{ZhangNiu2}%
  \BibitemOpen
  \bibfield  {author} {\bibinfo {author} {\bibfnamefont {L.}~\bibnamefont
  {Zhang}}\ and\ \bibinfo {author} {\bibfnamefont {Q.}~\bibnamefont {Niu}},\
  }\bibfield  {title} {\enquote {\bibinfo {title} {{Chiral Phonons at
  High-Symmetry Points in Monolayer Hexagonal Lattices}},}\ }\href@noop {}
  {\bibfield  {journal} {\bibinfo  {journal} {Phys. Rev. Lett.}\ }\textbf
  {\bibinfo {volume} {115}},\ \bibinfo {pages} {115502} (\bibinfo {year}
  {2015})}\BibitemShut {NoStop}%
\bibitem [{\citenamefont {Garanin}\ and\ \citenamefont
  {Chudnovsky}(2015)}]{Garanin}%
  \BibitemOpen
  \bibfield  {author} {\bibinfo {author} {\bibfnamefont {D.~A.}\ \bibnamefont
  {Garanin}}\ and\ \bibinfo {author} {\bibfnamefont {E.~M.}\ \bibnamefont
  {Chudnovsky}},\ }\bibfield  {title} {\enquote {\bibinfo {title} {{Angular
  momentum in spin-phonon processes}},}\ }\href@noop {} {\bibfield  {journal}
  {\bibinfo  {journal} {Phys. Rev. B}\ }\textbf {\bibinfo {volume} {92}},\
  \bibinfo {pages} {024421} (\bibinfo {year} {2015})}\BibitemShut {NoStop}%
\bibitem [{\citenamefont {Holanda}\ \emph {et~al.}(2018)\citenamefont
  {Holanda}, \citenamefont {Maior}, \citenamefont {Azevedo},\ and\
  \citenamefont {Rezende}}]{Holanda}%
  \BibitemOpen
  \bibfield  {author} {\bibinfo {author} {\bibfnamefont {J.}~\bibnamefont
  {Holanda}}, \bibinfo {author} {\bibfnamefont {D.~S.}\ \bibnamefont {Maior}},
  \bibinfo {author} {\bibfnamefont {A.}~\bibnamefont {Azevedo}}, \ and\
  \bibinfo {author} {\bibfnamefont {S.~M.}\ \bibnamefont {Rezende}},\
  }\bibfield  {title} {\enquote {\bibinfo {title} {{Detecting the phonon spin
  in magnon--phonon conversion experiments}},}\ }\href@noop {} {\bibfield
  {journal} {\bibinfo  {journal} {Nat. Phys.,
  http://dx.doi.org/10.1038/s41567-018-0079-y}\ } (\bibinfo {year}
  {2018})}\BibitemShut {NoStop}%
\bibitem [{\citenamefont {Kim}\ \emph {et~al.}(2017)\citenamefont {Kim} \emph
  {et~al.}}]{SKim}%
  \BibitemOpen
  \bibfield  {author} {\bibinfo {author} {\bibfnamefont {S.}~\bibnamefont
  {Kim}} \emph {et~al.},\ }\bibfield  {title} {\enquote {\bibinfo {title}
  {{Dynamically induced robust phonon transport and chiral cooling in an
  optomechanical system}},}\ }\href@noop {} {\bibfield  {journal} {\bibinfo
  {journal} {Nat. Commun.}\ }\textbf {\bibinfo {volume} {8}},\ \bibinfo {pages}
  {205} (\bibinfo {year} {2017})}\BibitemShut {NoStop}%
\bibitem [{\citenamefont {Zhu}\ \emph {et~al.}(2018)\citenamefont {Zhu} \emph
  {et~al.}}]{HZhu}%
  \BibitemOpen
  \bibfield  {author} {\bibinfo {author} {\bibfnamefont {H.}~\bibnamefont
  {Zhu}} \emph {et~al.},\ }\bibfield  {title} {\enquote {\bibinfo {title}
  {{Observation of chiral phonons}},}\ }\href@noop {} {\bibfield  {journal}
  {\bibinfo  {journal} {Science}\ }\textbf {\bibinfo {volume} {359}},\ \bibinfo
  {pages} {579} (\bibinfo {year} {2018})}\BibitemShut {NoStop}%
\bibitem [{\citenamefont {Jackson}(1998)}]{Jackson}%
  \BibitemOpen
  \bibfield  {author} {\bibinfo {author} {\bibfnamefont {J.~D.}\ \bibnamefont
  {Jackson}},\ }\href@noop {} {\emph {\bibinfo {title} {{Classical
  Electrodynamics}}}}\ (\bibinfo  {publisher} {Wiley \& Sons},\ \bibinfo
  {address} {New York},\ \bibinfo {year} {1998})\BibitemShut {NoStop}%
\bibitem [{\citenamefont {Gurevich}\ \emph {et~al.}(1975)\citenamefont
  {Gurevich}, \citenamefont {Klimov}, \citenamefont {Maiorov}, \citenamefont
  {Meleshko}, \citenamefont {Nikol'skii}, \citenamefont {Selivanov},\ and\
  \citenamefont {Suetin}}]{Gurevich}%
  \BibitemOpen
  \bibfield  {author} {\bibinfo {author} {\bibfnamefont {I.~I.}\ \bibnamefont
  {Gurevich}}, \bibinfo {author} {\bibfnamefont {A.~I.}\ \bibnamefont
  {Klimov}}, \bibinfo {author} {\bibfnamefont {V.~N.}\ \bibnamefont {Maiorov}},
  \bibinfo {author} {\bibfnamefont {E.~A.}\ \bibnamefont {Meleshko}}, \bibinfo
  {author} {\bibfnamefont {B.~A.}\ \bibnamefont {Nikol'skii}}, \bibinfo
  {author} {\bibfnamefont {V.~I.}\ \bibnamefont {Selivanov}}, \ and\ \bibinfo
  {author} {\bibfnamefont {V.~A.}\ \bibnamefont {Suetin}},\ }\bibfield  {title}
  {\enquote {\bibinfo {title} {{Magnetic field at a $\mu^{+}$ meson in a
  ferromagnet}},}\ }\href@noop {} {\bibfield  {journal} {\bibinfo  {journal}
  {Sov. Phys. $-$JETP}\ }\textbf {\bibinfo {volume} {42}},\ \bibinfo {pages}
  {222} (\bibinfo {year} {1975})}\BibitemShut {NoStop}%
\bibitem [{\citenamefont {Berger}\ and\ \citenamefont
  {Vroomen}(1965)}]{Berger}%
  \BibitemOpen
  \bibfield  {author} {\bibinfo {author} {\bibfnamefont {L.}~\bibnamefont
  {Berger}}\ and\ \bibinfo {author} {\bibfnamefont {A.~R.~De}\ \bibnamefont
  {Vroomen}},\ }\bibfield  {title} {\enquote {\bibinfo {title} {{Influence of
  the Internal Field on the Residual Resistance of Very Pure Iron}},}\
  }\href@noop {} {\bibfield  {journal} {\bibinfo  {journal} {J. Appl. Phys.}\
  }\textbf {\bibinfo {volume} {36}},\ \bibinfo {pages} {2777} (\bibinfo {year}
  {1965})}\BibitemShut {NoStop}%
\bibitem [{\citenamefont {Anderson}\ and\ \citenamefont
  {Gold}(1963)}]{Anderson}%
  \BibitemOpen
  \bibfield  {author} {\bibinfo {author} {\bibfnamefont {J.~R.}\ \bibnamefont
  {Anderson}}\ and\ \bibinfo {author} {\bibfnamefont {A.~V.}\ \bibnamefont
  {Gold}},\ }\bibfield  {title} {\enquote {\bibinfo {title} {{de Haas-van
  Alphen Effect and Internal Field in Iron}},}\ }\href@noop {} {\bibfield
  {journal} {\bibinfo  {journal} {Phys. Rev. Lett.}\ }\textbf {\bibinfo
  {volume} {10}},\ \bibinfo {pages} {227} (\bibinfo {year} {1963})}\BibitemShut
  {NoStop}%
\bibitem [{\citenamefont {Joseph}\ and\ \citenamefont
  {Thorsen}(1963)}]{Joseph}%
  \BibitemOpen
  \bibfield  {author} {\bibinfo {author} {\bibfnamefont {A.~S.}\ \bibnamefont
  {Joseph}}\ and\ \bibinfo {author} {\bibfnamefont {A.~C.}\ \bibnamefont
  {Thorsen}},\ }\bibfield  {title} {\enquote {\bibinfo {title} {{de Haas-van
  Alphen Effect and Fermi Surface in Nickel}},}\ }\href@noop {} {\bibfield
  {journal} {\bibinfo  {journal} {Phys. Rev. Lett.}\ }\textbf {\bibinfo
  {volume} {11}},\ \bibinfo {pages} {554} (\bibinfo {year} {1963})}\BibitemShut
  {NoStop}%
\bibitem [{\citenamefont {Hubbard}(1963)}]{Hubbard}%
  \BibitemOpen
  \bibfield  {author} {\bibinfo {author} {\bibfnamefont {J.}~\bibnamefont
  {Hubbard}},\ }\bibfield  {title} {\enquote {\bibinfo {title} {{Electron
  Correlations in Narrow Energy Bands}},}\ }\href@noop {} {\bibfield  {journal}
  {\bibinfo  {journal} {Proc. R. Soc. Lond. A}\ }\textbf {\bibinfo {volume}
  {276}},\ \bibinfo {pages} {238} (\bibinfo {year} {1963})}\BibitemShut
  {NoStop}%
\bibitem [{\citenamefont {Izuyama}(1960)}]{Izuyama2}%
  \BibitemOpen
  \bibfield  {author} {\bibinfo {author} {\bibfnamefont {T.}~\bibnamefont
  {Izuyama}},\ }\bibfield  {title} {\enquote {\bibinfo {title} {{Collective
  Excitations of Electrons in Degenerate Bands. I. Spin Waves in Stoner's Model
  of Ferromagnetism}},}\ }\href@noop {} {\bibfield  {journal} {\bibinfo
  {journal} {Prog. Theor. Phys.}\ }\textbf {\bibinfo {volume} {23}},\ \bibinfo
  {pages} {969} (\bibinfo {year} {1960})}\BibitemShut {NoStop}%
\bibitem [{\citenamefont {Yosida}(1996)}]{YosidaB}%
  \BibitemOpen
  \bibfield  {author} {\bibinfo {author} {\bibfnamefont {K.}~\bibnamefont
  {Yosida}},\ }\href@noop {} {\emph {\bibinfo {title} {{Theory of
  Magnetism}}}}\ (\bibinfo  {publisher} {Springer-Verlag},\ \bibinfo {address}
  {Heidelberg},\ \bibinfo {year} {1996})\BibitemShut {NoStop}%
\bibitem [{\citenamefont {Levine}(1962)}]{ADlevine}%
  \BibitemOpen
  \bibfield  {author} {\bibinfo {author} {\bibfnamefont {A.~D.}\ \bibnamefont
  {Levine}},\ }\bibfield  {title} {\enquote {\bibinfo {title} {{A Note
  Concerning the Spin of the Phonon}},}\ }\href@noop {} {\bibfield  {journal}
  {\bibinfo  {journal} {Nuovo Cimento}\ }\textbf {\bibinfo {volume} {26}},\
  \bibinfo {pages} {190} (\bibinfo {year} {1962})}\BibitemShut {NoStop}%
\bibitem [{\citenamefont {Holz}(1972)}]{Holz}%
  \BibitemOpen
  \bibfield  {author} {\bibinfo {author} {\bibfnamefont {A.}~\bibnamefont
  {Holz}},\ }\bibfield  {title} {\enquote {\bibinfo {title} {{Phonons in a
  strong static magnetic field}},}\ }\href@noop {} {\bibfield  {journal}
  {\bibinfo  {journal} {II Nuovo Cimento B}\ }\textbf {\bibinfo {volume} {9}},\
  \bibinfo {pages} {83} (\bibinfo {year} {1972})}\BibitemShut {NoStop}%
\bibitem [{\citenamefont {Solano-Carrillo}\ and\ \citenamefont
  {Millis}(2016)}]{Solano2}%
  \BibitemOpen
  \bibfield  {author} {\bibinfo {author} {\bibfnamefont {E.}~\bibnamefont
  {Solano-Carrillo}}\ and\ \bibinfo {author} {\bibfnamefont {A.~J.}\
  \bibnamefont {Millis}},\ }\bibfield  {title} {\enquote {\bibinfo {title}
  {{Theory of entropy production in quantum many-body systems}},}\ }\href@noop
  {} {\bibfield  {journal} {\bibinfo  {journal} {Phys. Rev. B}\ }\textbf
  {\bibinfo {volume} {93}},\ \bibinfo {pages} {224305} (\bibinfo {year}
  {2016})}\BibitemShut {NoStop}%
\bibitem [{\citenamefont {Yosida}(1957)}]{Yosida}%
  \BibitemOpen
  \bibfield  {author} {\bibinfo {author} {\bibfnamefont {K.}~\bibnamefont
  {Yosida}},\ }\bibfield  {title} {\enquote {\bibinfo {title} {{Magnetic
  Properties of Cu-Mn Alloys}},}\ }\href@noop {} {\bibfield  {journal}
  {\bibinfo  {journal} {Phys. Rev.}\ }\textbf {\bibinfo {volume} {106}},\
  \bibinfo {pages} {893} (\bibinfo {year} {1957})}\BibitemShut {NoStop}%
\bibitem [{\citenamefont {Mott}(1936)}]{Mott3}%
  \BibitemOpen
  \bibfield  {author} {\bibinfo {author} {\bibfnamefont {N.~F.}\ \bibnamefont
  {Mott}},\ }\bibfield  {title} {\enquote {\bibinfo {title} {{The Electrical
  Conductivity of Transition Metals}},}\ }\href@noop {} {\bibfield  {journal}
  {\bibinfo  {journal} {Proc. R. Soc. Lond. A}\ }\textbf {\bibinfo {volume}
  {153}},\ \bibinfo {pages} {699} (\bibinfo {year} {1936})}\BibitemShut
  {NoStop}%
\bibitem [{\citenamefont {Semenenko}\ and\ \citenamefont
  {Sudovtsov}(1962)}]{Semenenko}%
  \BibitemOpen
  \bibfield  {author} {\bibinfo {author} {\bibfnamefont {E.~E.}\ \bibnamefont
  {Semenenko}}\ and\ \bibinfo {author} {\bibfnamefont {A.~I.}\ \bibnamefont
  {Sudovtsov}},\ }\bibfield  {title} {\enquote {\bibinfo {title} {{Some
  Features in the Temperature Dependence of the Electrical Resistance of
  Ferromagnetic Metals at Low Temperatures}},}\ }\href@noop {} {\bibfield
  {journal} {\bibinfo  {journal} {Soviet Phys. $-$JETP}\ }\textbf {\bibinfo
  {volume} {15}},\ \bibinfo {pages} {708} (\bibinfo {year} {1962})}\BibitemShut
  {NoStop}%
\bibitem [{\citenamefont {Mott}(1935)}]{Mott}%
  \BibitemOpen
  \bibfield  {author} {\bibinfo {author} {\bibfnamefont {N.~F.}\ \bibnamefont
  {Mott}},\ }\bibfield  {title} {\enquote {\bibinfo {title} {{A Discussion of
  the Transition Metals on the Basis of Quantum Mechanics}},}\ }\href@noop {}
  {\bibfield  {journal} {\bibinfo  {journal} {Proc. Phys. Soc.}\ }\textbf
  {\bibinfo {volume} {47}},\ \bibinfo {pages} {571} (\bibinfo {year}
  {1935})}\BibitemShut {NoStop}%
\bibitem [{\citenamefont {Lide}(2003)}]{Lide}%
  \BibitemOpen
  \bibfield  {author} {\bibinfo {author} {\bibfnamefont {D.~R.}\ \bibnamefont
  {Lide}},\ }\href@noop {} {\emph {\bibinfo {title} {{ CRC Handbook of
  Chemistry and Physics}}}},\ \bibinfo {edition} {84th}\ ed.\ (\bibinfo
  {publisher} {CRC Press. Boca Raton},\ \bibinfo {address} {Florida},\ \bibinfo
  {year} {2003})\BibitemShut {NoStop}%
\bibitem [{\citenamefont {Fuchs}(1936)}]{Fuchs}%
  \BibitemOpen
  \bibfield  {author} {\bibinfo {author} {\bibfnamefont {K.}~\bibnamefont
  {Fuchs}},\ }\bibfield  {title} {\enquote {\bibinfo {title} {{A Quantum
  Mechanical Calculation of the Elastic Constants of Monovalent Metals}},}\
  }\href@noop {} {\bibfield  {journal} {\bibinfo  {journal} {Proc. R. Soc.
  Lond. A}\ }\textbf {\bibinfo {volume} {153}},\ \bibinfo {pages} {622}
  (\bibinfo {year} {1936})}\BibitemShut {NoStop}%
\bibitem [{\citenamefont {Nautiyal}\ and\ \citenamefont
  {Auluck}(1986)}]{Nautiyal}%
  \BibitemOpen
  \bibfield  {author} {\bibinfo {author} {\bibfnamefont {T.}~\bibnamefont
  {Nautiyal}}\ and\ \bibinfo {author} {\bibfnamefont {S.}~\bibnamefont
  {Auluck}},\ }\bibfield  {title} {\enquote {\bibinfo {title} {{Electronic
  structure of ferromagnetic iron: Band structure and optical properties}},}\
  }\href@noop {} {\bibfield  {journal} {\bibinfo  {journal} {Phys. Rev. B}\
  }\textbf {\bibinfo {volume} {34}},\ \bibinfo {pages} {2299} (\bibinfo {year}
  {1986})}\BibitemShut {NoStop}%
\bibitem [{\citenamefont {Singal}\ and\ \citenamefont {Das}(1977)}]{Singal}%
  \BibitemOpen
  \bibfield  {author} {\bibinfo {author} {\bibfnamefont {C.~M.}\ \bibnamefont
  {Singal}}\ and\ \bibinfo {author} {\bibfnamefont {T.~P.}\ \bibnamefont
  {Das}},\ }\bibfield  {title} {\enquote {\bibinfo {title} {{Electronic
  structure of ferromagnetic hcp cobalt. I. Band properties}},}\ }\href@noop {}
  {\bibfield  {journal} {\bibinfo  {journal} {Phys. Rev. B}\ }\textbf {\bibinfo
  {volume} {16}},\ \bibinfo {pages} {5068} (\bibinfo {year}
  {1977})}\BibitemShut {NoStop}%
\bibitem [{\citenamefont {Wang}\ and\ \citenamefont {Callaway}(1974)}]{Wang}%
  \BibitemOpen
  \bibfield  {author} {\bibinfo {author} {\bibfnamefont {C.~S.}\ \bibnamefont
  {Wang}}\ and\ \bibinfo {author} {\bibfnamefont {J.}~\bibnamefont
  {Callaway}},\ }\bibfield  {title} {\enquote {\bibinfo {title} {{Band
  structure of nickel: Spin-orbit coupling, the Fermi surface, and the optical
  conductivity}},}\ }\href@noop {} {\bibfield  {journal} {\bibinfo  {journal}
  {Phys. Rev. B}\ }\textbf {\bibinfo {volume} {9}},\ \bibinfo {pages} {4897}
  (\bibinfo {year} {1974})}\BibitemShut {NoStop}%
\bibitem [{\citenamefont {Kittel}(1996)}]{Kittel}%
  \BibitemOpen
  \bibfield  {author} {\bibinfo {author} {\bibfnamefont {C.}~\bibnamefont
  {Kittel}},\ }\href@noop {} {\emph {\bibinfo {title} {{Introduction to Solid
  State Physics}}}},\ \bibinfo {edition} {7th}\ ed.\ (\bibinfo  {publisher}
  {Wiley \& Sons},\ \bibinfo {address} {New York},\ \bibinfo {year}
  {1996})\BibitemShut {NoStop}%
\bibitem [{\citenamefont {Watson}\ and\ \citenamefont
  {Bennett}(1977)}]{Watson}%
  \BibitemOpen
  \bibfield  {author} {\bibinfo {author} {\bibfnamefont {R.~E.}\ \bibnamefont
  {Watson}}\ and\ \bibinfo {author} {\bibfnamefont {L.~H.}\ \bibnamefont
  {Bennett}},\ }\bibfield  {title} {\enquote {\bibinfo {title} {{Calculation of
  atomic hyperfine-field coupling constants}},}\ }\href@noop {} {\bibfield
  {journal} {\bibinfo  {journal} {Phys. Rev. B}\ }\textbf {\bibinfo {volume}
  {15}},\ \bibinfo {pages} {502} (\bibinfo {year} {1977})}\BibitemShut
  {NoStop}%
\bibitem [{\citenamefont {Townes}\ \emph {et~al.}(1950)\citenamefont {Townes},
  \citenamefont {Herring},\ and\ \citenamefont {Knight}}]{Townes}%
  \BibitemOpen
  \bibfield  {author} {\bibinfo {author} {\bibfnamefont {C.~H.}\ \bibnamefont
  {Townes}}, \bibinfo {author} {\bibfnamefont {C.}~\bibnamefont {Herring}}, \
  and\ \bibinfo {author} {\bibfnamefont {W.~D.}\ \bibnamefont {Knight}},\
  }\bibfield  {title} {\enquote {\bibinfo {title} {{The Effect of Electronic
  Paramagnetism on Nuclear Magnetic Resonance Frequencies in Metals}},}\
  }\href@noop {} {\bibfield  {journal} {\bibinfo  {journal} {Phys. Rev.}\
  }\textbf {\bibinfo {volume} {77}},\ \bibinfo {pages} {852} (\bibinfo {year}
  {1950})}\BibitemShut {NoStop}%
\bibitem [{\citenamefont {Song}\ \emph {et~al.}(1972)\citenamefont {Song},
  \citenamefont {Trooster},\ and\ \citenamefont {Benczer-Koller}}]{Song}%
  \BibitemOpen
  \bibfield  {author} {\bibinfo {author} {\bibfnamefont {C.-J.}\ \bibnamefont
  {Song}}, \bibinfo {author} {\bibfnamefont {J.}~\bibnamefont {Trooster}}, \
  and\ \bibinfo {author} {\bibfnamefont {N.}~\bibnamefont {Benczer-Koller}},\
  }\bibfield  {title} {\enquote {\bibinfo {title} {{Measurement of 2$s$ and
  3$s$ Electron Spin Density in Iron Metal}},}\ }\href@noop {} {\bibfield
  {journal} {\bibinfo  {journal} {Phys. Rev. Lett.}\ }\textbf {\bibinfo
  {volume} {29}},\ \bibinfo {pages} {1165} (\bibinfo {year}
  {1972})}\BibitemShut {NoStop}%
\bibitem [{\citenamefont {Duff}\ and\ \citenamefont {Das}(1975)}]{Duff2}%
  \BibitemOpen
  \bibfield  {author} {\bibinfo {author} {\bibfnamefont {K.~J.}\ \bibnamefont
  {Duff}}\ and\ \bibinfo {author} {\bibfnamefont {T.~P.}\ \bibnamefont {Das}},\
  }\bibfield  {title} {\enquote {\bibinfo {title} {{Hyperfine field in metallic
  iron: An appraisal of the theory}},}\ }\href@noop {} {\bibfield  {journal}
  {\bibinfo  {journal} {Phys. Rev. B}\ }\textbf {\bibinfo {volume} {12}},\
  \bibinfo {pages} {3870} (\bibinfo {year} {1975})}\BibitemShut {NoStop}%
\bibitem [{\citenamefont {Bennett}\ \emph {et~al.}(1970)\citenamefont
  {Bennett}, \citenamefont {Watson},\ and\ \citenamefont {Carter}}]{Bennett2}%
  \BibitemOpen
  \bibfield  {author} {\bibinfo {author} {\bibfnamefont {L.~H.}\ \bibnamefont
  {Bennett}}, \bibinfo {author} {\bibfnamefont {R.~E.}\ \bibnamefont {Watson}},
  \ and\ \bibinfo {author} {\bibfnamefont {G.~C.}\ \bibnamefont {Carter}},\
  }\bibfield  {title} {\enquote {\bibinfo {title} {{Relevance of Knight Shift
  Measurements to the Electronic Density of States}},}\ }\href@noop {}
  {\bibfield  {journal} {\bibinfo  {journal} {J. Res. Natl. Bur. Stand. A}\
  }\textbf {\bibinfo {volume} {74}},\ \bibinfo {pages} {569} (\bibinfo {year}
  {1970})}\BibitemShut {NoStop}%
\bibitem [{\citenamefont {Muto}\ \emph {et~al.}(1965)\citenamefont {Muto},
  \citenamefont {Kobayasi},\ and\ \citenamefont {Hayakawa}}]{Muto}%
  \BibitemOpen
  \bibfield  {author} {\bibinfo {author} {\bibfnamefont {T.}~\bibnamefont
  {Muto}}, \bibinfo {author} {\bibfnamefont {S.}~\bibnamefont {Kobayasi}}, \
  and\ \bibinfo {author} {\bibfnamefont {H.}~\bibnamefont {Hayakawa}},\
  }\bibfield  {title} {\enquote {\bibinfo {title} {{On Indirect Knight Shift
  and NMR in Ferrromagnetic Metals Part III. Numerical Calculation of NMR
  Frequency and Hyperfine Field in Ni, Co and Fe.}}}\ }\href@noop {} {\bibfield
   {journal} {\bibinfo  {journal} {J. Phys. Soc. Jpn.}\ }\textbf {\bibinfo
  {volume} {20}},\ \bibinfo {pages} {1167} (\bibinfo {year}
  {1965})}\BibitemShut {NoStop}%
\bibitem [{\citenamefont {Gold}(1974)}]{Gold}%
  \BibitemOpen
  \bibfield  {author} {\bibinfo {author} {\bibfnamefont {A.~V.}\ \bibnamefont
  {Gold}},\ }\bibfield  {title} {\enquote {\bibinfo {title} {{Review paper:
  Fermi surfaces of the ferromagnetic transition metals}},}\ }\href@noop {}
  {\bibfield  {journal} {\bibinfo  {journal} {J. Low Temp. Phys.}\ }\textbf
  {\bibinfo {volume} {16}},\ \bibinfo {pages} {3} (\bibinfo {year}
  {1974})}\BibitemShut {NoStop}%
\bibitem [{\citenamefont {Stearns}(1973)}]{Stearns}%
  \BibitemOpen
  \bibfield  {author} {\bibinfo {author} {\bibfnamefont {M~.B.}\ \bibnamefont
  {Stearns}},\ }\bibfield  {title} {\enquote {\bibinfo {title} {{On the Origin
  of Ferromagnetism and the Hyperfine Fields in Fe, Co, and Ni}},}\ }\href@noop
  {} {\bibfield  {journal} {\bibinfo  {journal} {Phys. Rev. B}\ }\textbf
  {\bibinfo {volume} {8}},\ \bibinfo {pages} {4383} (\bibinfo {year}
  {1973})}\BibitemShut {NoStop}%
\bibitem [{\citenamefont {Bloembergen}(1950)}]{Bloembergen2}%
  \BibitemOpen
  \bibfield  {author} {\bibinfo {author} {\bibfnamefont {N.}~\bibnamefont
  {Bloembergen}},\ }\bibfield  {title} {\enquote {\bibinfo {title} {{On the
  Ferromagnetic Resonance in Nickel and Supermalloy}},}\ }\href@noop {}
  {\bibfield  {journal} {\bibinfo  {journal} {Phys. Rev.}\ }\textbf {\bibinfo
  {volume} {78}},\ \bibinfo {pages} {572} (\bibinfo {year} {1950})}\BibitemShut
  {NoStop}%
\bibitem [{\citenamefont {Holmes}\ and\ \citenamefont
  {Alexandrakis}(1977)}]{Holmes}%
  \BibitemOpen
  \bibfield  {author} {\bibinfo {author} {\bibfnamefont {J.~B.}\ \bibnamefont
  {Holmes}}\ and\ \bibinfo {author} {\bibfnamefont {G.~C.}\ \bibnamefont
  {Alexandrakis}},\ }\bibfield  {title} {\enquote {\bibinfo {title}
  {{Observation of nonlinear phenomena in ferromagnetic transmission
  resonance}},}\ }\href@noop {} {\bibfield  {journal} {\bibinfo  {journal}
  {Phys. Rev. B}\ }\textbf {\bibinfo {volume} {16}},\ \bibinfo {pages} {484}
  (\bibinfo {year} {1977})}\BibitemShut {NoStop}%
\bibitem [{\citenamefont {Agranat}\ \emph {et~al.}(1984)\citenamefont
  {Agranat}, \citenamefont {Ashitkov}, \citenamefont {Granovskii},\ and\
  \citenamefont {Rukman}}]{Agranat}%
  \BibitemOpen
  \bibfield  {author} {\bibinfo {author} {\bibfnamefont {M.~B.}\ \bibnamefont
  {Agranat}}, \bibinfo {author} {\bibfnamefont {S.~I.}\ \bibnamefont
  {Ashitkov}}, \bibinfo {author} {\bibfnamefont {A.~B.}\ \bibnamefont
  {Granovskii}}, \ and\ \bibinfo {author} {\bibfnamefont {G.~I.}\ \bibnamefont
  {Rukman}},\ }\bibfield  {title} {\enquote {\bibinfo {title} {{Interaction of
  picosecond laser pulses with the electron, spin, and phonon subsystems of
  nickel}},}\ }\href@noop {} {\bibfield  {journal} {\bibinfo  {journal} {Soviet
  Phys. $-$JETP}\ }\textbf {\bibinfo {volume} {59}},\ \bibinfo {pages} {804}
  (\bibinfo {year} {1984})}\BibitemShut {NoStop}%
\bibitem [{\citenamefont {Vaterlaus}\ \emph {et~al.}(1990)\citenamefont
  {Vaterlaus}, \citenamefont {Guarisco}, \citenamefont {Lutz}, \citenamefont
  {Aeschlimann}, \citenamefont {Stampanoni},\ and\ \citenamefont
  {Meir}}]{Vaterlaus}%
  \BibitemOpen
  \bibfield  {author} {\bibinfo {author} {\bibfnamefont {A.}~\bibnamefont
  {Vaterlaus}}, \bibinfo {author} {\bibfnamefont {D.}~\bibnamefont {Guarisco}},
  \bibinfo {author} {\bibfnamefont {M.}~\bibnamefont {Lutz}}, \bibinfo {author}
  {\bibfnamefont {M.}~\bibnamefont {Aeschlimann}}, \bibinfo {author}
  {\bibfnamefont {M.}~\bibnamefont {Stampanoni}}, \ and\ \bibinfo {author}
  {\bibfnamefont {F.}~\bibnamefont {Meir}},\ }\bibfield  {title} {\enquote
  {\bibinfo {title} {{Different spin and lattice temperatures observed by
  spin-polarized photoemission with picosecond laser pulses}},}\ }\href@noop {}
  {\bibfield  {journal} {\bibinfo  {journal} {J. Appl. Phys.}\ }\textbf
  {\bibinfo {volume} {67}},\ \bibinfo {pages} {5661} (\bibinfo {year}
  {1990})}\BibitemShut {NoStop}%
\end{thebibliography}%

\end{document}